\documentclass[journal]{IEEEtran}
\ifCLASSINFOpdf
\else
\fi

\usepackage{xcolor}
\usepackage{graphics}
\usepackage{graphicx}
\usepackage{subfig}

\usepackage{lipsum}
\makeatletter  
\def\@IEEEBIOphotowidth{1.5cm}    
\def\@IEEEBIOphotodepth{1.5cm}   
\def\@IEEEBIOhangwidth{1.7cm}    
\def\@IEEEBIOhangdepth{1.7cm}    

\usepackage{soul}

\usepackage{etoolbox}
\newtoggle{finalPaper}

\toggletrue{finalPaper} 

\iftoggle{finalPaper} {

	\newcommand{\rmvtxt}[1]{}}
{
	
	\newcommand{\rmvtxt}[1]{\st{#1}}}

\begin{document}
%
\title{Trust-as-a-Service: A reputation-enabled\\trust framework for 5G networks}
%
%
%

\author{Jos\'e Mar\'ia Jorquera Valero, Pedro Miguel S\'anchez S\'anchez, Manuel Gil P\'erez,\\ Alberto Huertas Celdr\'an, and Gregorio Mart\'inez P\'erez,~\IEEEmembership{Member,~IEEE}

\thanks{Jos\'e Mar\'ia Jorquera Valero, Pedro Miguel S\'anchez S\'anchez, Manuel Gil P\'erez, and Gregorio Mart\'inez P\'erez are with the Department of Information and Communications Engineering, University of Murcia, 30100 Murcia, Spain (e-mail: josemaria.jorquera@um.es; pedromiguel.sanchez@um.es; mgilperez@um.es; gregorio@um.es) \textit{(Corresponding author: Jos\'e Mar\'ia Jorquera Valero)}}
\thanks{Alberto Huertas Celdr\'an is with the Communication Systems Group (CSG) at the Department of Informatics (IfI), University of Zurich UZH, 8050 Zürich, Switzerland (e-mail: huertas@ifi.uzh.ch).}}

%
%

\markboth{Journal of \LaTeX\ Class Files,~Vol.~14, No.~8, August~2015}%
{Shell \MakeLowercase{\textit{et al.}}: Bare Demo of IEEEtran.cls for IEEE Journals}
%



\maketitle

\begin{abstract}

Trust, security, and privacy are three of the major pillars to assemble the fifth generation network and beyond. Despite such pillars are principally interconnected, they arise a multitude of challenges to be addressed separately. 5G ought to offer flexible and pervasive computing capabilities across multiple domains according to user demands and assuring trustworthy network providers. Distributed marketplaces expect to boost the trading of heterogeneous resources so as to enable the establishment of pervasive service chains between cross-domains. Nevertheless, the need for reliable parties as ``marketplace operators'' plays a pivotal role to achieving a trustworthy ecosystem. One of the principal blockages in managing foreseeable networks is the need of adapting previous trust models to accomplish the new network and business requirements. In this regard, this article is centered on trust management of 5G multi-party networks. The design of a reputation-based trust framework is proposed as a Trust-as-a-Service (TaaS) solution for any distributed multi-stakeholder environment where zero trust and zero-touch principles should be met. Besides, a literature review is also conducted to recognize the network and business requirements currently envisaged. Finally, the validation of the proposed trust framework is performed in a real research environment, the 5GBarcelona testbed, leveraging 12\% of a 2.1GHz CPU with 20 cores and 2\% of the 30GiB memory. In this regard, these outcomes reveal the feasibility of the TaaS solution in the context of determining reliable network operators.


\end{abstract}



%
\IEEEpeerreviewmaketitle


\section{Introduction}

Among all pillars of the communication evolution, fifth generation technologies play a paramount role as cutting-edge network paradigms, from radio technology and optical networks to non-terrestrial network communications and ubiquitous computing. Such paradigms in turn bring challenges to be overcome by next-generation technologies such as reduction of energy footprint, multi-tenancy, extreme-reliable communication, automated management and orchestration, on-demand service and resource allocation, and trustworthy infrastructures, among others \cite{silva20205g}.

In 5G and beyond, the distributed marketplaces encompass a potential solution commonly utilized by the vertical industry to cater for end-to-end composite services or slices that allow satisfying all requirements and Key Performance Indicators (KPIs) in terms of coverage, networking and computing resources, and Virtual Network Functions (VNFs). Since such heterogeneous services and resources may be supplied by a single provider or multi-party collaboration across several domains, it is paramount to elect a trustworthy network provider, which ensure the fulfillment of requirements and KPIs, and guarantee a trustworthy environment. In this regard, trust models facilitate reliable establishments among different stakeholders predicting a forthcoming trust score.

Nevertheless, trust models need to progress over time as novel network and business requirements are constantly appearing and they cannot be covered by prior trust models \cite{JORQUERAVALERO2022103596}. 5G and beyond are envisioned as compounded networks in which an end-to-end communication will entail multiple entities from the same or different locations and domains. Thus, trust models ought to contemplate reliable end-to-end chains to predict future behaviors of all implicated entities from the origin to the end. In the same manner, implicit trust should not be granted to stakeholders, regardless of whether they are placed in an intra- or inter-domain scenario, as trust by default is a potential attack vector exploited by spiteful entities. In this sense, a zero trust approach, driven by the NIST \cite{stafford2020zero}, is a predominant principle for imminent trust models to dwindle the attack surface. Another fundamental requirement is the minimization of human interaction in the trust model lifecycle management, also known as zero-touch approaches. Trust models should spur the automatization of network and service management via high-level policies, triggers, and artificial intelligence algorithms. Simultaneously, the automation process also entails an essential effort to enable easier integration with other B5G network orchestration and management components; for instance, a distributed marketplace allowing verticals to expose telco digital assets and hire them to automatically satisfy user demands. Nevertheless, these requirements are currently not all addressed at the same time by the solutions in the literature \cite{JORQUERAVALERO2022103596}.

Hence, the paper at hand analyzes the present literature to determine whether the identified network and business requirements related to trust models are being contemplated. Besides, it also presents a reputation-based trust framework capable of guaranteeing a trustworthy ecosystem where stakeholders can establish reliable end-to-end connections across domains as well as dealing with the above-mentioned novel network requirements. In particular, such a framework considers a set of product offers (POs), available in the 5GZORRO European project distributed marketplace \cite{fernandez2021multi}, to be thoroughly analyzed. Thereupon, an adapted PeerTrust model for peer-to-peer communities is leveraged to predict both a provider and product offer trust scores from historical interactions and recommendations, the latter published in a Data Lake platform to be consulted by interested stakeholders.

The remainder of this article can be outlined as follows. Section \ref{sec:related_work} carries out in-depth research into the utmost importance trust models applied to on-demand service and resource provision environments. Section \ref{sec:framework_design} describes the design of our reputation-based trust framework. Then, Section \ref{sec:experiments} presents the performance assessment results of the trust framework. Finally, Section \ref{sec:conclusions} recaps some conclusions as well as ongoing works for future work.


\section{Related Work}
\label{sec:related_work}

In this section, we analyze the most newfangled approaches that explore trust models as a mechanism to cater for reliable on-demand service or resource capabilities in 5G and B5G. To determine the compatibility level of analyzed approaches, we have compiled a set of universal network and business requirements which should be shared between 5G trust model solutions such as trustworthy end-to-end chains across domains and the zero trust and zero-touch principles.

Trust remains a vital requirement in the cloud environment since reputable relationships between consumers and providers may guarantee the fulfillment of offered user's Quality of Service (QoS) as well as dwindling the chance to infringing Service Level Agreements (SLAs) or Smart Contracts (SCs) signed. In addition, the current cloud environments share paramount characteristics with the on-demand service and resource capability provisioning, in which our reputation-based trust framework is entailed. For these reasons, we have identified a set of researches that not only contemplate trust models as a potential solution for cloud environments but also fit to some extent with the requirements described above.

Hassan et al. proposed in \cite{hassan2020enhanced} a QoS-based model to assess the trust of cloud providers dynamically before each new interaction. The authors composed the cloud resources' trust as a blend of the provider reputation, from users' feedback, plus the computing power at run-time, from SLA attributes. To regulate dishonest feedback, the covariance technique was leveraged to calculate user credibility and discover misleading feedback. Nevertheless, their model did not consider requirements such as end-to-end chains (only analyzing the last entity of the chain) or the zero trust principle (main ideas are not supported through paper). To test their enhanced QoS-based model, the authors contrasted their transaction success rate (TSR) against the Armor model, obtaining 0.92 and 0.74 TSRs, respectively, when there was a 40\% of fake users' feedback. 

Concerning the business needs to scale their computing and infrastructure capabilities up, a new term called Federated Cloud appeared to enable the integration of public, community, and private clouds to support business requirements. Thereby, an inappropriate selection of deployment cloud platforms may encompass performance, security, and even legal issues. To cope with them, Papadakis et al. presented in \cite{papadakis2019collaborative} a hybrid reputation-based trust management (RTM) approach to evade selecting untrustworthy cloud applications in federated scenarios. The authors leveraged KPIs and Quality of Experience (QoE) metrics to continuously evaluate cloud providers, according to customer deployment objectives and credibility. In this case, the biased assessments were detected by the deviation of objective SLA monitoring values against a customer's rating. Additionally, the authors fulfilled pivotal requirements such as automatization, by considering trust as a substep of the full automated SLA lifecycle, and trustworthy end-to-end connections, by analyzing reputation of both cloud infrastructure and applications. The experiments showcased a 20\% performance betterment thanks to the credibility mechanism. In the same line, Latif et al. presented in \cite{latif2021novel} a federated cloud trust management framework to insure the fulfillment of privacy laws and the protection of customer's data. As a result, they addressed the issues for trust establishment and evaluation. Particularly, the framework was formed of three dimensions: SLA parameters focused on security and privacy, feedback from customers, and feedback from other clouds. Since the final reputation is composed from multiple entities involved in the relationship, it can be ensured end-to-end trustworthy chains. To test their framework, the authors contrasted their trust scores against other existing schemes and their SLAs such as IBM, Amazon, or Google, among others, reaching the second-best result in the vast majority of cloud providers. Nonetheless, they did not describe ideas aligned the zero trust principle so it cannot be guaranteed.

Also dealing with the trust in cloud environments, in \cite{khilar2019trust}, Khilar et al. centered on ascertaining both customers' trust prior to accessing the cloud and resources' trust. This approach was formed by multiple sub-modules among which the available resource catalog can be highlighted as the starting point. Like in \cite{papadakis2019collaborative} and in \cite{latif2021novel}, customers and resources' behavior, feedback, and SLA parameters were contemplated to compute a trust score. As part of SLA parameters, the customer's satisfaction was formulated as the total number of successful tasks. Furthermore, the authors ensured zero trust, full automation of all steps, and an assessment of multiple entities involved in the end-to-end chain, not only the final target. In terms of performance, the k-Nearest Neighbors (kNN) got a 0.94 of precision, recall, and F1-Score as well as a 6.48\% of mean absolute error (MAE), and 25.45\% of Root MAE. Not only the selection of trustworthy Cloud Providers (CPs) is critical but also the possible co-tenants hosted in the cloud. In \cite{thakur2019robust}, Takur and Breslin proposed a reputation-based management mechanism utilized by CPs to distinguish users' behavior and properly assign resources based on trust scores. The CP reputation was formulated as the capability and willingness to differentiate between good and malicious users. Hence, a constant increase or decrease of users' reputation of a multi-tenancy group enhanced the CP reputation; otherwise, the CP was not able to create homogeneous groups and its reputation dwindled. In addition, the authors considered feasible rational, irrational, and opportunistic reports by the CP, achieving a higher reputation when the rational approach was met.

Another indispensable requirement to be fulfilled by 5G and beyond trust models is the establishment of a reliable end-to-end chain. In this vein, Wang et al. proposed in \cite{wang2021mobile} a trust evaluation model for mobile edge nodes (TEM-MENs) to guarantee a reliable node chain between the trustor and the trustee, withstanding malicious attacks. Depending on the number of nodes, the authors declared an atomic trust chain (without intermediate nodes) or a combined trust chain. In the case of the atomic chain, they considered interaction trust, energy trust, and recommendation trust, along with time windows. With regard to the combined chain, they collected the previous values for each node forming the path(s), which covers the end-to-end trustworthy chain requirement. To automatically collect all information and fulfill the zero-touch philosophy, an enhanced Dijkstra algorithm was employed. By mean of several experiments, the TEM-MENs demonstrated to have the highest detection rate (96\%) and its runtime had a slope of 0.1, so it allowed better adaptability. Similarly, Fan et al. also dealt with Service Function Chains (SFC) in \cite{fan2021credibility}. In particular, they designed a credibility-based deployment strategy (CBDS) to prognosticate the trust of VNF nodes through the SFC credibility. The credibility was formed by reliability, availability, and authenticity properties. Additionally, the authors also added extra functionalities such as adopted sliding windows and trigger mechanisms, which empowered the mechanism as dynamic and fully automatic. In terms of experiments, the CBDS reached a 90\% acceptance rate for 0.75 credibility value (the highest one). 

Lastly, Debe et al. addressed in \cite{debe2019towards} the problem of ensuring trust in the provision of compute and networking capabilities. Hence, the authors presented a reputation-based solution to discover a trust score in a decentralized way. A blockchain together with SCs allowed calculating the reputation of public fog nodes from past interactions. Besides, the credibility of client IoT devices, which was computed as a clustering of the most legitimate group of vectors contrasting the rating rate with the majority, was also contemplated to build the final trust of each public fog node. Similarly, reward and punishment mechanisms were defined to regulate the weight of client feedback and detect potential malicious clients.

As observed, only \cite{khilar2019trust} meets the three principal requirements: the zero trust, zero-touch, and trustworthy end-to-end chains at the same time, which are paramount pillars to provide a reputation-based model compatible with future networks. Yet, this approach did not cover the same objective as ours because it was oriented to an access control cloud scenario and our objective is to enable a trustworthy resource and service provisioning discovery for distributed marketplaces. The proposed framework therefore aims to fill the gap in reputation-based trust models for 5G networks, as well as to ensure an automated, practical, and scalable framework.


\section{Reputation-based Trust Framework Design}
\label{sec:framework_design}

\begin{figure*}
    \centering
    \includegraphics[width=0.9\linewidth]{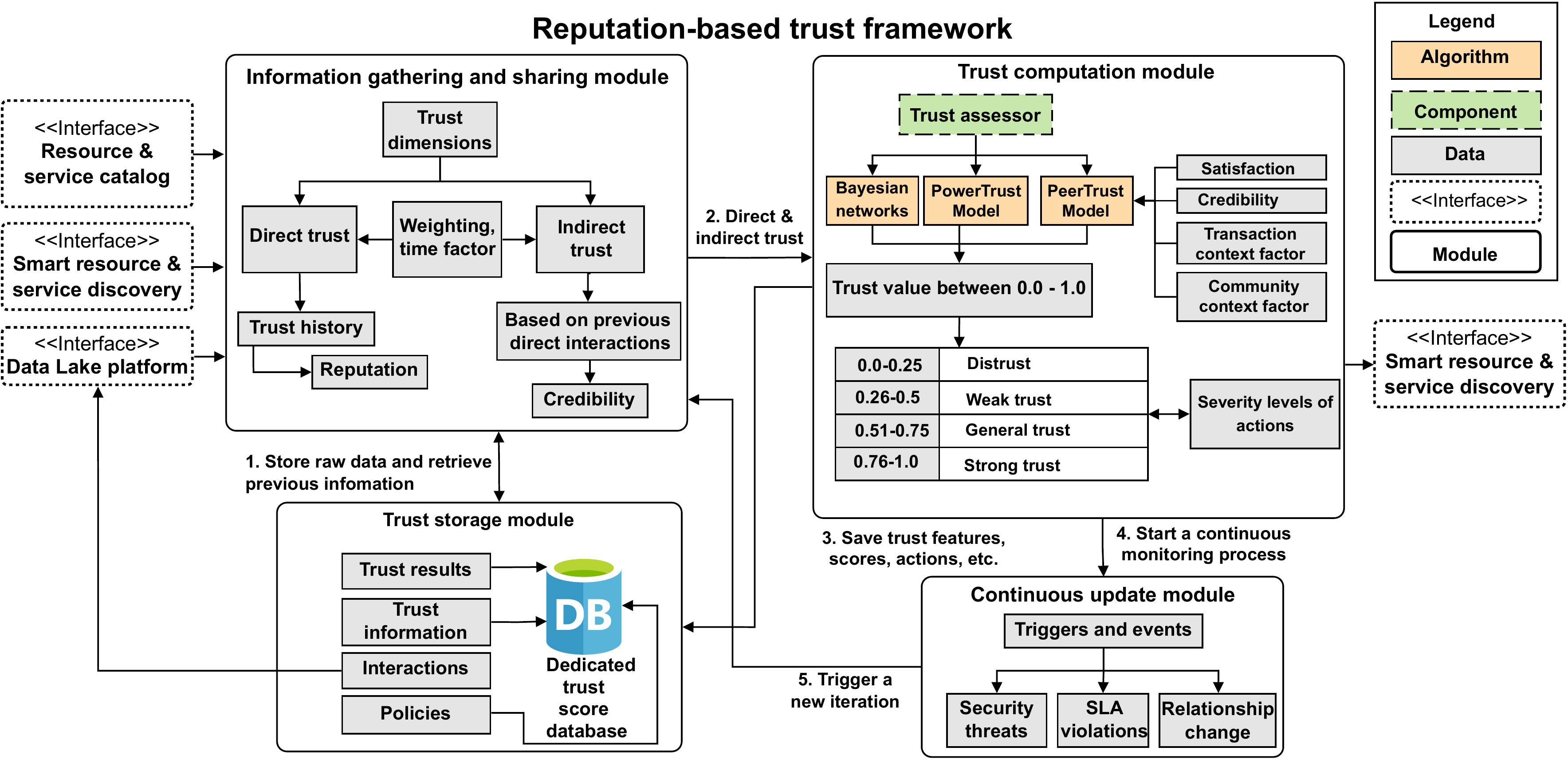}
    \caption{Design of the proposed reputation-based trust framework}
    \label{fig:reputation_trust_framework_design}
\end{figure*}

This section describes the principal modules and characteristics of our Trust-as-a-Service solution as well as the most important pillars to compute trust scores. As shown in \figurename~\ref{fig:reputation_trust_framework_design}, the trust framework is principally composed of four sub-modules: the \textit{Information gathering and sharing}, the \textit{Trust computation}, the \textit{Trust storage}, and the \textit{Continuous update}. Next subsections thoroughly explain the utmost important steps under each module. By means of modules, we will contextualize how zero trust, zero-touch, and reliable end-to-end chain principles can be addressed. Note that such a trust and reputation framework has been designed under the 5GZORRO project \cite{carrozo2020ai}, and in consequence, a few concepts will be briefly introduced throughout the following subsections to contextualize and detail the decisions taken.

\subsection{Information gathering and sharing module}

First and foremost, the trust framework collects raw data from multiple information sources such as the Data Lake, and resource and service catalog (see \figurename~\ref{fig:reputation_trust_framework_design}). To contextualize, the Data Lake is a centralized and shared data environment in its native format that leverages big data and the catalog refers to a decentralized repository employed for identifying available resource and service. These information sources provision an extremely fruitful range of data for our trust and reputation framework. It can infer data like SLA breach predictions related to certain resource and service providers and statistic parameters with respect to the available product offers in the distributed marketplace platform, etc.

Depending on how the information is gathered, the TaaS classifies it into direct and indirect trust. When it comes to direct trust, it is linked to personal experiences the trustor had after interacting with a trustee. In particular, our proposal is centered on the reputation deemed through providers and offer trust histories. Concerning indirect trust, it is conventionally collected from third-party recommendations. Yet, recommendations do not always come from trustworthy third-parties. The proposed trust framework integrates two mechanisms to avoid misleading recommendations as much as possible. Firstly, a feedback credibility mechanism is leveraged to figure out the recommender honesty through the Personalized Similarity Metric (PSM). On another hand, our TaaS also integrates a dynamic list of trustworthy recommenders, which is originated from previously computed trust scores. The list is continuously updated after each new interaction and it also contemplates the time factor to weigh up the most up-to-date scores as the most relevant. Lastly, it should be highlighted that the information gathering steps should be applied to each entity involved in the trust chain and not only to the extremes.

\subsection{Trust computation module}
\label{sec:trust_computation_module}

Once all available trust information has been collected, such information is forwarded to the trust computation module to be processed. It is worth mentioning that this step is carried out regardless of whether stakeholders have a previous trust relationship for some time or whether they belong to the same domain, an intra-domain relationship, since the zero trust principle must be complied with. In spite of the reputation-based trust framework expects to support multiple trust assessors such as Bayesian networks and the PowerTrust model in the foreseeable future (see \figurename~\ref{fig:reputation_trust_framework_design}), this manuscript is principally centered on the PeerTrust model \cite{xiong2004peertrust}. PeerTrust is based on interactions and designed for distributed environments in which satisfaction, feedback credibility, transaction factor, and community factor are the pillars to build trustworthy establishments. Nevertheless, it only introduces the theoretical concepts under the above-mentioned pillars described in \cite{xiong2004peertrust} and not how they should be formulated. Because of that, we decided to go for the PeerTrust model as it brings a high flexibility level to adapt the trust model to the final enforcement scenario as well as meet the distributed philosophy followed by the 5GZORRO marketplace platform, in which this framework is being utilized. In this vein, we have designed and developed several pillars to delineate the final trust scores. Yet, since the stakeholder's satisfaction is the utmost importance pillar and the most complex, we have put a special emphasis on it.

Firstly, the satisfaction represented below in (\ref{eq:satisfaction}) measures the acceptance degree that a stakeholder $u$ has with another after finishing the \textit{i-th trustworthy interaction}. In this regard, we considered two key dimensions to discover the final satisfaction value, the provider's satisfaction and the product offer's satisfaction. Note that $\psi$ and $\phi$ are the weights of each dimension and they must satisfy that $\psi + \phi = 1$. 

\begin{equation}
    S(u,i) = \psi \cdot PS(u,i) + \phi \cdot PO(u,I)
    \label{eq:satisfaction}
\end{equation}

Concerning the provider's satisfaction, it is composed of three main features, as shown in (\ref{eq:provider_satisfaction}): the reputation of a stakeholder $j$ (\textit{Rep}); a set of $n$ recommendations (\textit{Rec}) about the target $j$ from a trusted third party (TTP); and the last trust score for each recommender in the previous set. Thence, the satisfaction of stakeholder $u$ on the \textit{i-th} interaction will be computed about the target stakeholder $j$.

\begin{equation}
    PS(u,j) = Rep(u,j)\cdot\left(\displaystyle\bigoplus_{x=1}^{n} Rec(x,j)\cdot T^{(t-1)}(u,x)\right)
    \label{eq:provider_satisfaction}
\end{equation}


In this sense, \textit{Rep} represents the average reputation that the stakeholder $u$ has on the stakeholder $j$ contemplating all its available resources and services. This reputation function calculated below in (\ref{eq:provider_reputation}) considers features such as available assets (AA), total assets (IA), available assets at a given location (AAL), total assets at a given location (IAL), the total number of predicted SLA violations (PV) that were lastly managed both successful (MV ) and unsuccessful (EV ), and no-predicted SLA violations (NPV). In addition, multiple time windows are also deemed together with weighting factors to cater for higher relevance to the newest scores.

\begin{equation}
    \resizebox{.99\hsize}{!}{$Rep(u,j) = \displaystyle\sum_{k=1}^{n} \varepsilon(k)\cdot \frac{\bigg( \frac{AA(j)} {IA(j)} + \frac {AAL(j)} {IAL(j)} + 2\cdot  \frac {MV(j)} {PV(j)} - 2\cdot \frac {EV(j)+NPV(j)} {PV(j)} \bigg) +2}{6}$}
    \label{eq:provider_reputation}
\end{equation}

When it comes to the provider's satisfaction, we additionally leverage an aggregation operation based on the arithmetic mean to combine recommendations ($Rec$) about the stakeholder $j$ with the last trust score provided by the recommender. Once the computation of a provider's satisfaction has been completed, the next step is to reckon the product offer's satisfaction. To deal with it, the reputation-based trust framework utilizes similar dimensions but adapted to only take into account information about a particular product offer since the provider's satisfaction considers all available assets. In the 5GZORRO ecosystem, there are five types of product offers defined in the resource and service catalog: radio access network (RAN), spectrum, VNF/container network function (CNF), slice, and edge. Hence, the product offer's satisfaction is formed by reputation (\textit{Rep}), recommendations (\textit{Rec}), and last trust scores of a specific type of offer and provider.

With respect to credibility, our approach follows the PSM as it may be applied to multiple contexts. In particular, such a metric determines how similar two unfamiliar stakeholders are when evaluating a set of stakeholders. Thereby, the similarity is the metric leveraged to contrast the opinions about a target stakeholder as well as measuring the distance of credibility about a set of stakeholders assessed by both stakeholders. Hence, the higher credibility distance after evaluating the same set of stakeholder, the less credible the opinion. Additionally, the reputation-based trust framework introduces two context factors that allow adjusting the final trust score to the current transaction type and the community see \figurename~\ref{fig:reputation_trust_framework_design}. Firstly, the transaction context factor pillar intends to forecast a trust value linked to the current interaction, with a particular stakeholder or product offer, from the number of feedbacks published in the Data Lake from different time windows. The transaction context factor rewards stakeholders who publish their interactions with others in the Data Lake since it spurs future stakeholders to look into the Data Lake, request recommendations to other stakeholders, and grow the community. Finally, the community context factor pillar attempts to gather multiple feedback about a target stakeholder from a dynamic list of trustworthy recommenders. Hence, the community context factor measures the number of interactions that a specific stakeholder had in the community through the contribution of services or resources with other stakeholders. Besides, multiple recommendations together with the credibility of the recommender are contemplated through an aggregation function to achieve the general reputation of the community about a target stakeholder.

In the end, the weighting of the credibility, satisfaction, and transaction context factor plus the community context factor enables to determine a final trust score of a target stakeholder by contemplating multiple interactions and reputations from different recommenders and time windows.

\subsection{Trust storage module}

After computing a trust score, the next step is to save both the raw and inferred data for future establishments and recommendations. To cope with it, the proposed framework manages two types of information storage sources (see \figurename~\ref{fig:reputation_trust_framework_design}). Because the TaaS is instantiated per domain, a private non-relational database has been contemplated per instance. The dedicated database stores principally information with respect to raw data collected from information sources such as the Data Lake and resource and service catalog, the PeerTrust model information, and lastly, the trust scores. Furthermore, it may also save internal policies or rules to be used by the continuous update module so as to trigger events or detect misbehaviors. Thus, sensitive information will be stored in the dedicated database as it is not going to be shared with other stakeholders. In addition to the private database, the framework also leverages a Data Lake. In particular, the capital aim is to spread knowledge about trustworthy interactions among stakeholders that form the 5GZORRO ecosystem. In this regard, newcomers may recognize feasible recommenders to be consulted. In the same manner, other stakeholders who already established previous trust interactions across domains may request new recommendations when their trustworthy recommenders are not able to support feedback. In contrast to the private database, the Data Lake cannot store sensitive information since its objective is to be consulted by countless stakeholders. However, the Data Lake introduces key features such as a long-term reputation reflection and traceability.

\subsection{Continuous update module}

Parallel to the trust storage module, and once a trust score has been concluded, the trust computation module triggers a continuous update process focused on the target stakeholder. This module plays a pivotal role as it may enable earlier identification of plausible attacks through a suitable configuration of triggers, events, and rules. In this sense, the continuous update module not only ameliorates the security capabilities of the framework via context-dependence and dynamicity but also empowers an end-to-end automatization. Therefore, should unfamiliar phenomenons appear in ongoing trust relationships, the reputation-based trust framework can identify them and take the proper decisions. 

With regard to real-time events in an established trust relationship, the continuous update module presents reward and punishments mechanisms to oversee the stakeholders' behaviors. Hence, a trust score is recalculated after appearing new events. Among the principal events contemplated to increase or dwindle trust scores, we consider security threats, change policy relationships, SLA violations, service execution failures, and decay of time, to name but a few. Therefore, whether negative events occur, the previous trust scores will be diminished applying the proper internal policies, and in consequence, it could be finished to discover more reliable stakeholders. On the contrary, favorable events entail an increase in the previous trust score. Note that the reward and punishment mechanisms also support the zero-touch principle as it enables to fully automation of the proposed framework. Thence, human interaction is not required and the reputation-bases trust framework can smoothly interact with other modules participating in the resource and service provisioning discovery, as can be observed below.


\section{Experiments and Results}
\label{sec:experiments}

This section introduces the principal characteristics related to the 5GBarcelona testbed in terms of CPU and memory. Additionally, we showcase multiple experiments in which the CPU and memory consumption can be observed for different amount of product offers. Similarly, we can also analyze the required to run each module of the framework.

\subsection{Experimental setting}

On-demand resource and service allocation to cover user requirements is a real challenge beyond 5G networks. Therefore, the 5GZORRO project introduces an innovative solution through a distributed marketplace platform, which enables a secure and trustworthy provisioning of resource and service capabilities. In addition to the marketplace, there is a component named Smart Resource and Service Discovery (SRSD) that enforces zero-touch capabilities and allows obtaining a set of resources and services through an intent-based discovery. 
The SRSD and the trust framework have been integrated to ensure TaaS and elect the final provider based on its trust scores together with other intents. By means of such an integration, it is possible to expose reliable telco digital assets and hire them and enable a zero-touch interaction with other network orchestration and management components.

In terms of testbeds, the reputation-based trust framework has been deployed in the 5GBarcelona where the framework was instantiated in a 1vCPU of a 6th Intel Xeon Gold 5218R with 2.1 GHz and 20 cores. Particularly, the framework was deployed in a worker with 8 cores and 30 GiB.

\subsection{Performance Evaluation}

To check the proper functionality of the proposed framework, we performed several experiments.

\begin{figure}[!ht]
  \centering
  \vspace{-4mm}
  \subfloat[Average CPU consumption per offer]{\includegraphics[width=0.48\textwidth]{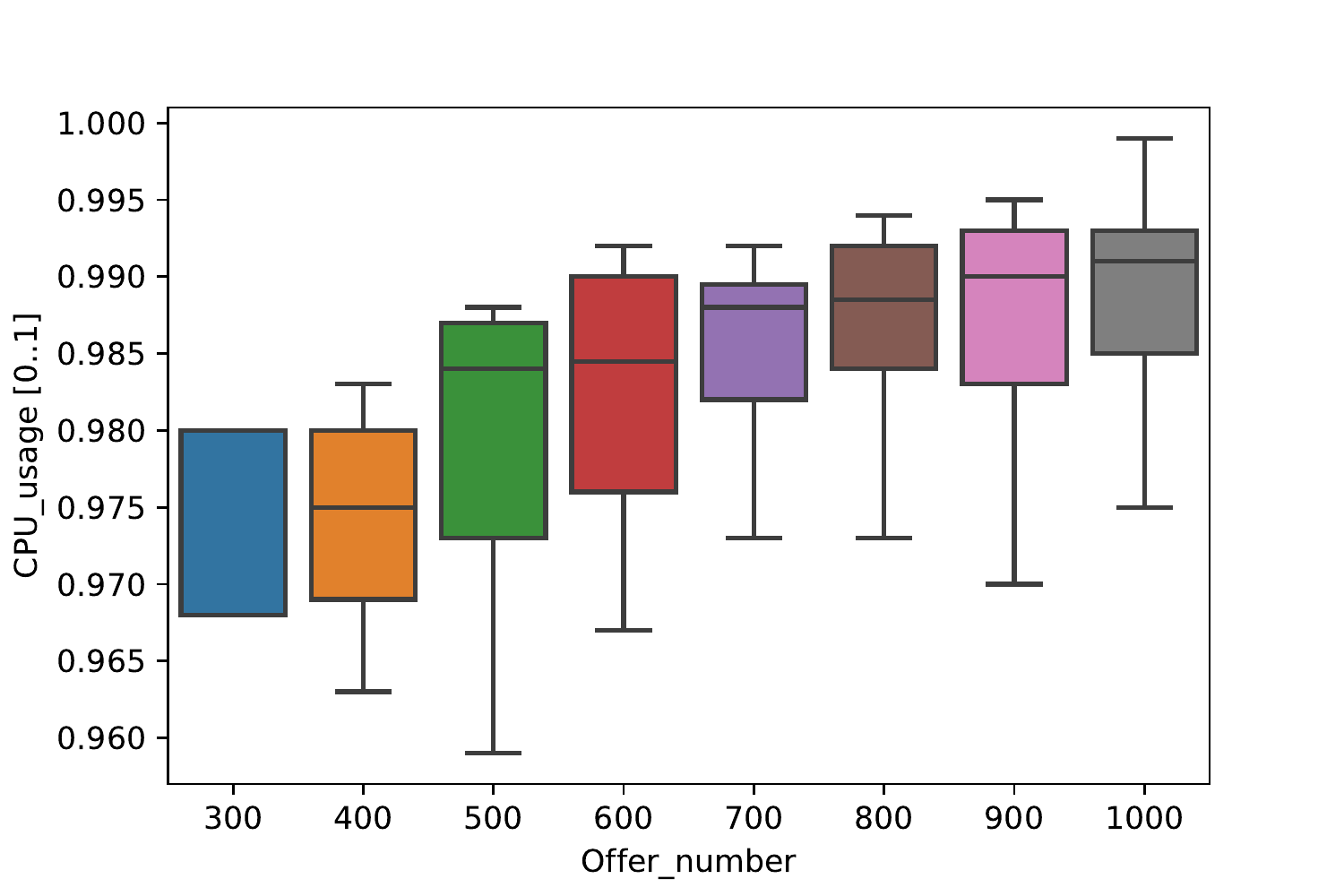}\label{fig:CPU}}
  \\
  \subfloat[Average memory consumption per offer]{\includegraphics[width=0.48\textwidth]{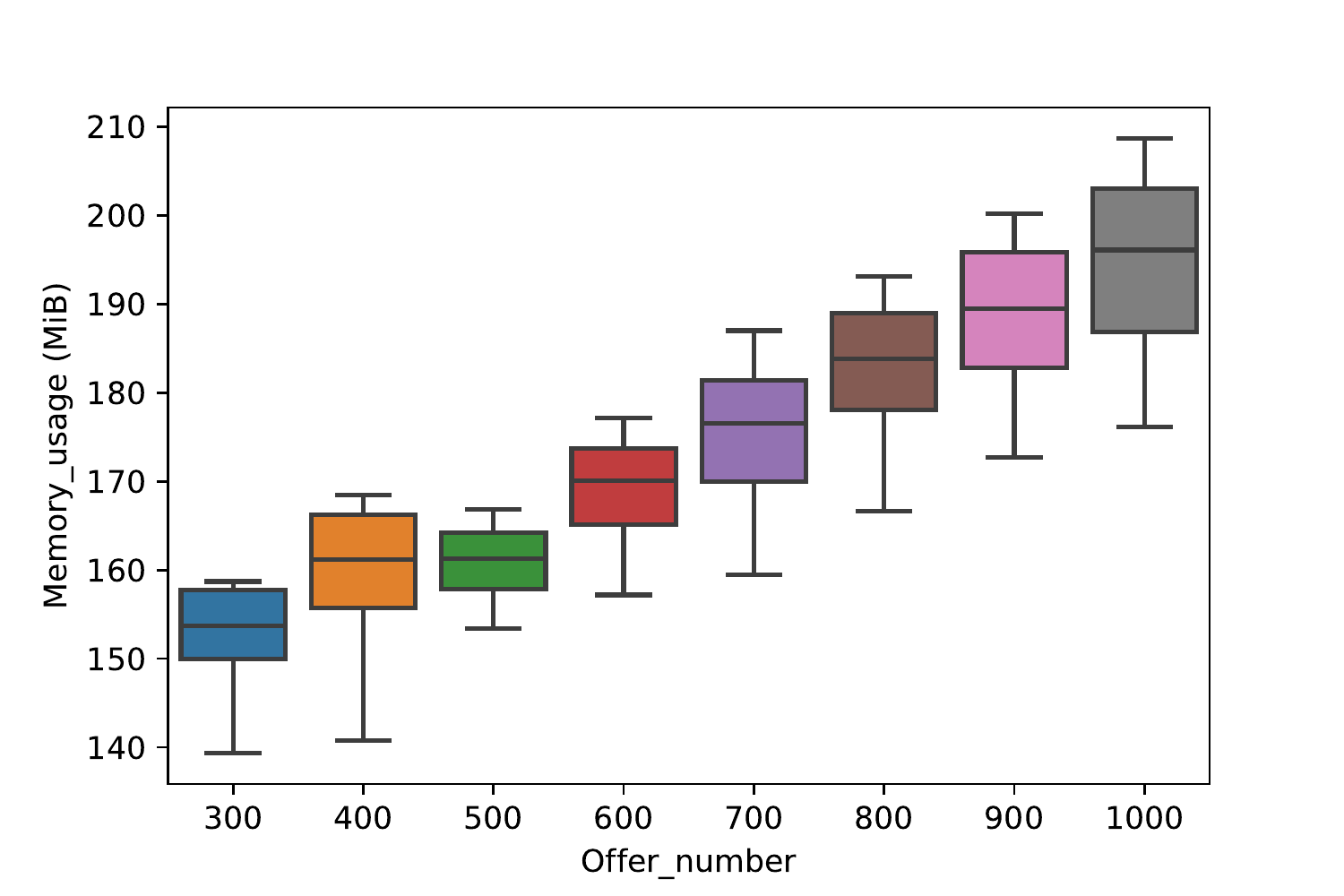}\label{fig:memory}}
  \caption{Reputation-based trust framework consumption}
  \label{fig:cpu_memory_consumption}
\end{figure}

Firstly, we analyzed the CPU and memory consumption required by our framework. Thereby, \figurename~\ref{fig:cpu_memory_consumption} displays an light growth when the number of offers to be processed increases. In particular, the average CPU consumption of the 1vCPU (only 1 core) allocated to the framework lies between 0.975 and 0.993 (see \figurename~\ref{fig:CPU}), except for the case of 100 and 200 offers in which are 0.477 and 0.483 respectively. In spite of our vCPU was being used almost 99\%, we only consumed 12\% of the total CPU available in the server. In the case of memory, the framework required around 155 and 195 MiB (only 2\% of the total memory) to process the multiple sets of product offers (see \figurename~\ref{fig:memory}), except for 100 and 200 offers in which the average was set to 140 and 152 MiB. The values related to 100 and 200 offers were not plotted since they might complicate the box plot visualization and the number of CPU and memory measurements was lower than the rest of offers because the framework computed trust scores quicker.

\begin{figure}[!t]
    \centering
    \includegraphics[width=0.48\textwidth]{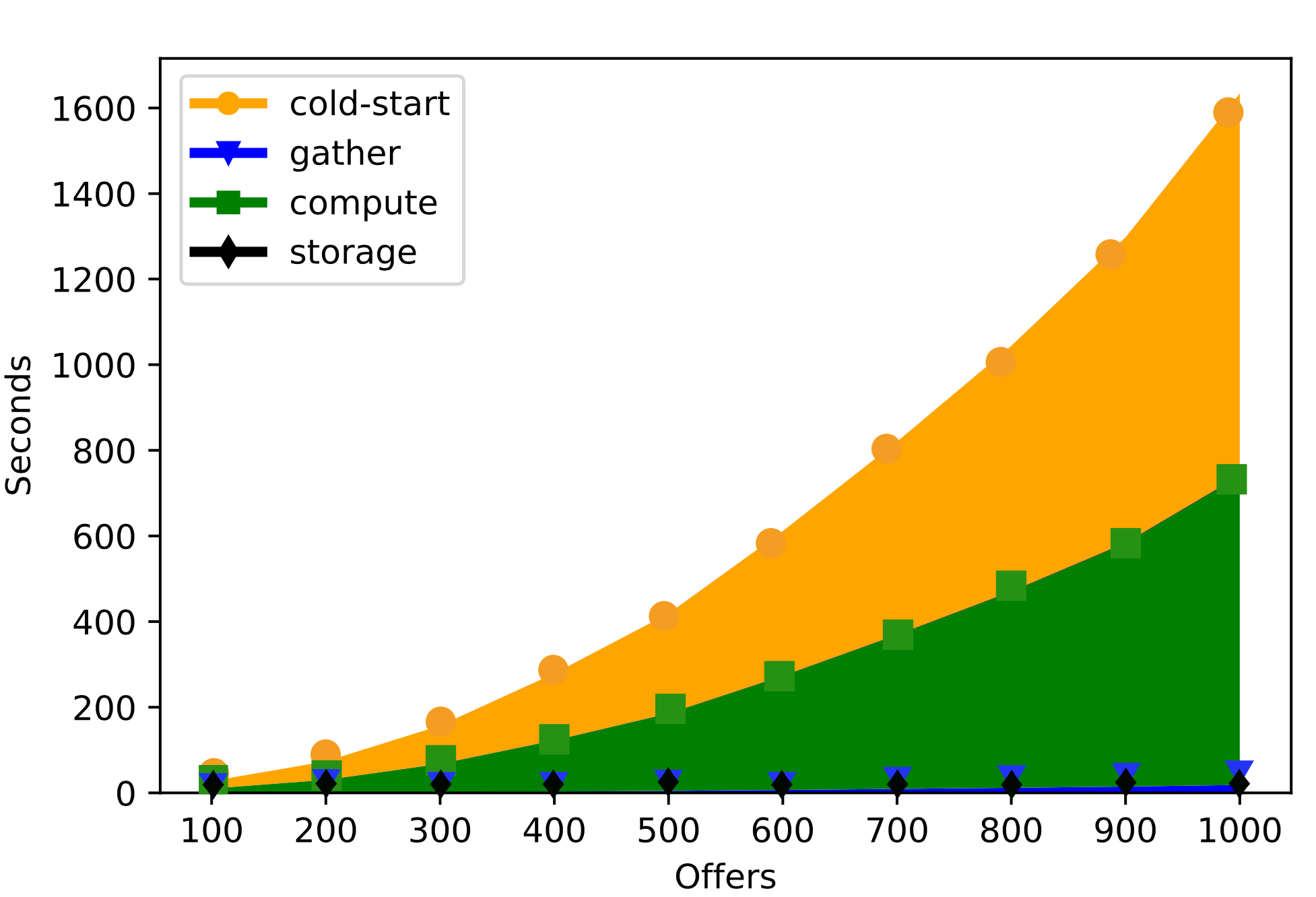}
    \caption{Reputation-based trust framework consumption time with cold-start}
    \label{fig:curve_time_offer}
    \vspace{-6mm}
\end{figure}

When it comes to consuming time, the cold-start mechanism doubled the required time to provide trust scores for each set of offers, as depicted in \figurename~\ref{fig:curve_time_offer}. Nevertheless, such a mechanism was leveraged because of 5GZORRO ecosystem is not fully instantiated and there is not information enough about trust relationships. In the foreseeable future, the cold-start mechanism will be eradicated and consequently its consumption time. In this regard, the proposed framework was able to perform the gathering information, computation, and storage phases in 10.4, 185.6, and 741.4 seconds for 100, 500, and 1000 offers, respectively, in a sequential way.

\begin{figure}[!h]
    \vspace{-4mm}
    \centering
    \includegraphics[width=0.53\textwidth]{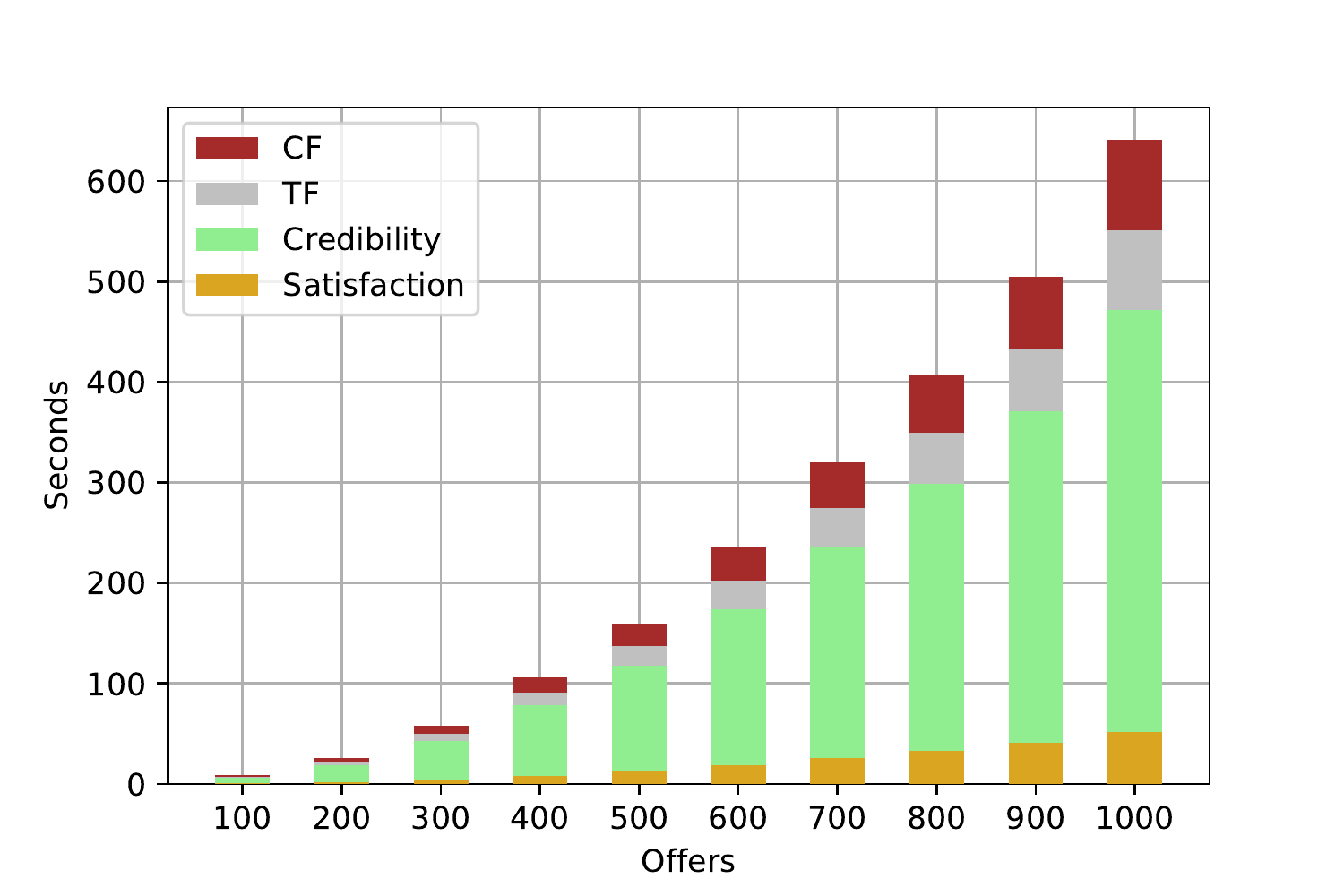}
    \caption{Time distribution by computation phase}
    \label{fig:curve_time_compute}
\end{figure}

Since the compute phase entailed the higher time, we decided to analyze it in detail. In \figurename~\ref{fig:curve_time_compute}, we can observe a light exponential increment in which the credibility consumed more than half of the total time to carry out the three phases aforementioned, 5.6 out of 10.4 seconds, 104.9 out of 185.6, and 419.9 out of 741.4 for 100, 500, and 1000 offers. Hence, a feasible approach to address this drawback would be to parallel not only the computation sub-phases but also the gathering and the storage phases. In this sense, we could execute different phases of multiple processes in parallel and intend to decrease the required time to provide final trust scores.

From previous experiments, it can be concluded that the reputation-based trust framework does not require a huge amount of CPU and RAM to request and analyze information as well as compute trust scores for a high number of offers, for instance, 500 POs. However, the necessary time to calculate trust scores should be decreased in next iterations since 180 seconds were consumed to evaluate the trust of 500 POs.  Despite of that, the authors consider the proposed framework meets with the expected behavior and performance as other filtering mechanisms, such as intent-based or hardware requirements, are applied before computing trust scores. Therefore, the PO number to be analyzed by the proposed framework is normally ranked from 200 and 500.


\section{Conclusion and Future Work}
\label{sec:conclusions}

The article at hand analyzes some of the most prominent trust and reputation approaches in the research field to identify the utilization of network and business requirements. Taking into account the previous investigations, we propose a reputation-based trust framework capable of helping stakeholders in the decision of electing the most reliable providers of resource or service capabilities available in a distributed marketplace. At the same time, the proposed framework takes into account critical requirements of 5G multi-party networks as aforementioned. 

In terms of computation, an adapted PeerTrust model has been selected as the most befitting algorithm to foresee end-to-end trust scores based on historical interactions and recommendations. Multiple experiments were carried out in the 5GBarcelona testbed. The outputs showed that the trust and reputation framework had a slight increase of consumption time when the number of offers was increased. However, this increase can be reduced by parallelism techniques.

As future work, we will extend the current functionalities of the framework to contemplate prominent algorithms such as Bayesian networks and the PowerTrust model and contrast them with PeerTrust to analyse their performances and accuracies. In this sense, the TaaS will also be deployed in the 5TONIC testbed, from Telefonica, to contrast metrics. Besides, additional functionality to cope with potential trust attacks such as collusion, Sybil, bad-mouthing, among others, should be designed and developed after looking at the ones that our trust framework may suffer. Finally, we will analyze feasible parallelism techniques to be applied in the framework to reduce the consumption time.


\section*{Acknowledgment}

This work has been supported by the European Commission through 5GZORRO project (grant no. 871533) part of the 5G PPP in Horizon 2020. 

\ifCLASSOPTIONcaptionsoff
  \newpage
\fi

\bibliographystyle{IEEEtran}
\bibliography{references}
\vspace{-16mm}
\begin{IEEEbiography}
   [{\includegraphics[height=1.8cm, width=1.45cm]{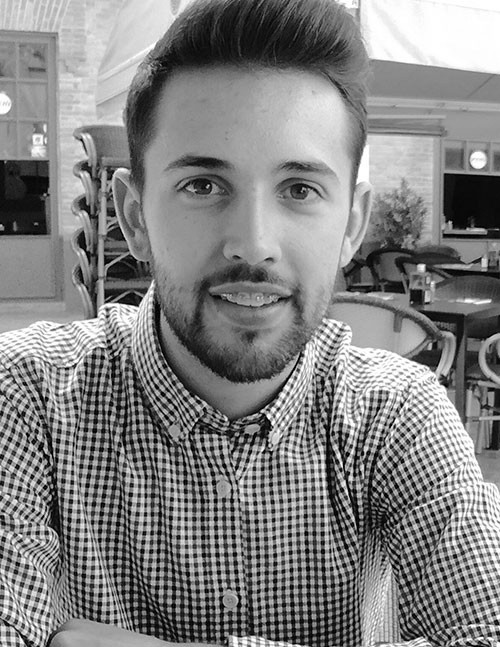}}]{Jos\'e Mar\'ia Jorquera Valero} is a Ph.D. student in Computer Science at Murcia University. Jorquera Valero received the M.Sc. degree in Computer Science from the University of Murcia, Spain. His scientific interests include trust, security, 5G, data privacy, continuous authentication, and cybersecurity.
\end{IEEEbiography}
\vspace{-20mm}
\begin{IEEEbiography}
   [{\includegraphics[height=1.8cm, width=1.5cm]{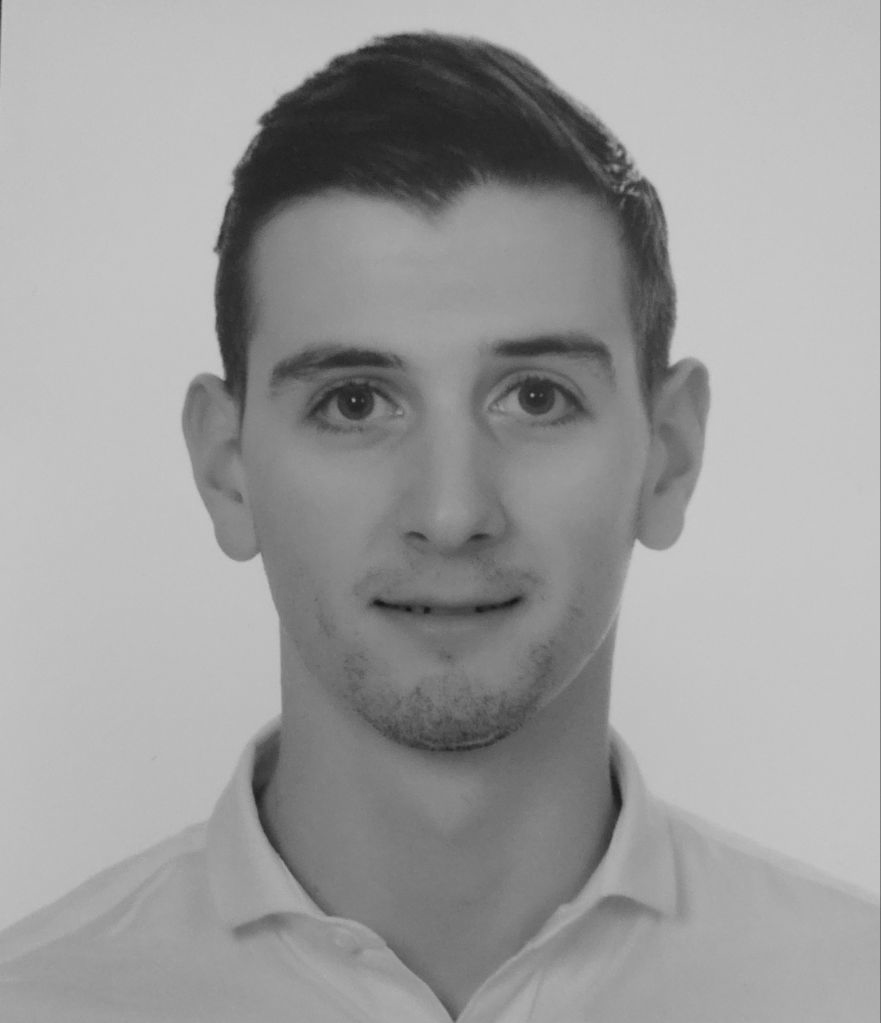}}]{Pedro Miguel S\'anchez S\'anchez} received the M.Sc. degree in computer science from the University of Murcia. He is currently pursuing his PhD in computer science at University of Murcia. His research interests are focused on continuous authentication, networks, 5G, cybersecurity and the application of machine learning and deep learning to the previous fields.
\end{IEEEbiography}
\vspace{-18mm}
\begin{IEEEbiography}
   [{\includegraphics[height=1.8cm, width=1.5cm]{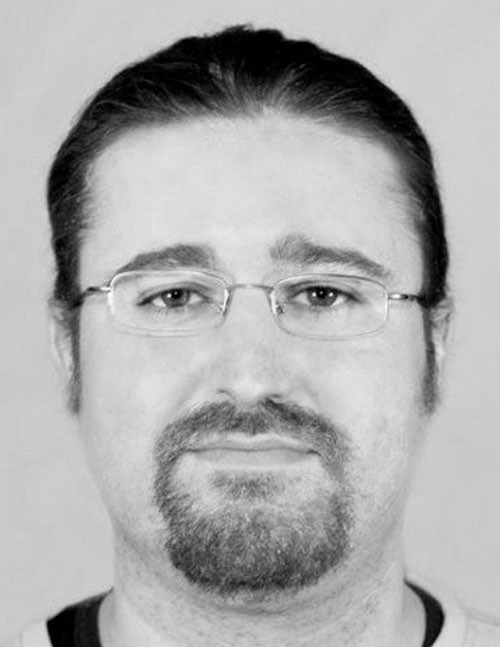}}]{Manuel Gil P\'erez} is Associate Professor in the Department of Information and Communication Engineering of the University of Murcia, Spain. His scientific activity is mostly focused on cybersecurity, including intrusion detection systems, trust and reputation management, and security operations in highly dynamic scenarios. He received M.Sc. and Ph.D. degrees (the latter with distinction) in Computer Science from the same University. 
\end{IEEEbiography}
\vspace{-18mm}
\begin{IEEEbiography}
   [{\includegraphics[height=1.8cm, width=1.5cm]{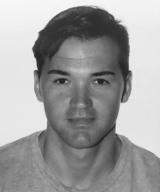}}]{Alberto Huertas Celdr\'an} received the M.Sc. and Ph.D. degrees in computer science from the University of Murcia, Spain. He is currently a postdoctoral fellow associated with the Communication Systems Group (CSG) at the University of Zurich UZH. His scientific interests include medical cyber-physical systems (MCPS), brain–computer interfaces (BCI), cybersecurity, data privacy, continuous authentication, semantic technology, context-aware systems, and computer networks.
\end{IEEEbiography}
\vspace{-18mm}
\begin{IEEEbiography}
   [{\includegraphics[height=1.8cm, width=1.6cm]{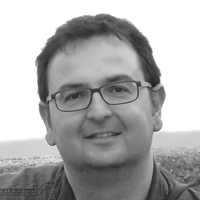}}]{Gregorio Mart\'inez P\'erez} is Full Professor in the Department of Information and Communications Engineering of the University of Murcia, Spain. His scientific activity is mainly devoted to cybersecurity and networking, also working on the design and autonomic monitoring of real-time and critical applications and systems. He is working on different national (14 in the last decade) and European IST research projects (11 in the last decade) related to these topics, being Principal Investigator in most of them. 
\end{IEEEbiography}



\end{document}